\documentclass[12pt]{iopart}

\usepackage{iopams}
\usepackage{graphicx}
\usepackage{algorithm2e}
\usepackage{epsfig}
\begin{document}

\title{Intra-City Urban Network and Traffic Flow Analysis from GPS Mobility Trace}

\author{Ian X. Y. Leung$^{1}$, Shu-Yan Chan$^1$, Pan Hui$^2$ and Pietro Li\`{o}$^1$}

\address{$^1$ Computer Laboratory, University of Cambridge, Cambridge CB3 0FD, U.K.}
\address{$^2$ Deutsche Telekom Laboratories, Ernst-Reuter-Platz 7, 10709 Berlin, Germany}
\ead{ixyl2@cl.cam.ac.uk}

\begin{abstract}
We analyse two large-scale intra-city urban networks and traffic flows therein measured by GPS traces of taxis in San Francisco and Shanghai. Our results coincide with previous findings that, based purely on topological means, it is often insufficient to characterise traffic flow. Traditional shortest-path betweenness analysis, where shortest paths are calculated from each pairs of nodes, carries an unrealistic implicit assumption that each node or junction in the urban network generates and attracts an equal amount of traffic. We also argue that weighting edges based only on Euclidean distance is inadequate, as primary roads are commonly favoured over secondary roads due to the perceived and actual travel time required. We show that betweenness traffic analysis can be improved by a simple extended framework which incorporates both the notions of node weights and fastest-path betweenness. We demonstrate that the framework is superior to traditional methods based solely on simple topological perspectives.
\end{abstract}

\maketitle

\section{Introduction}
Complex networks provide a natural abstraction of the structure and relationships between entities in real-world systems. From the study of social relationships to biological systems, the study of Network Science has provided abundant tools and research opportunities for the understanding and modelling of such complex systems. Urban networks, those pertaining to city infrastructures, have been traditionally subject to investigation from the fields of Urban Planning, Economic Geography, Economics, and Engineering. In a common node-link nomenclature, road junctions are considered as nodes while roads are treated as edges connecting the nodes in the network. Typical real-world networks from the realms of social and biological systems are famously known to exhibit properties such as scale-free degree distributions and the small world phenomenon, where despite their large size, the average distance between any pairs of node remain relatively small. On the other hand, urban networks are unlike the aforementioned networks due to spatial and geographical constraints. They are known to exhibit a smaller average degree and longer diameter, due to them being almost planar~\cite{Gastner2006, Cardillo2006}.

Studies from both Urban Planning and Network Science have revealed interesting correlations between topological properties of urban networks to human related activities in a wide variety of city topologies~\cite{CalabreseRatti2010, porta2009street}. We are interested in further understanding the relationship between topological properties and traffic flow with the support of large-scale mobility traces. Previous works have shed light into topological-based traffic prediction, but either focussed on unrealistic or incomplete traffic estimation or was done in a small and coarse scale. The development of cheap and portable Global Positioning System (GPS) has greatly advanced the state of the art of understanding human mobility. By mining large-scale and real-time tracking datasets, one is able empirically evaluate important hypotheses. Such tools have for instance eased communication network deployment, e.g. network operators can choose to deploy more network facilities such as Wi-Fi access points or relay nodes for mobile communication at areas with high traffic flow~\cite{Hui2010}.

The purpose of this paper is to analyse various network-based metrics and their ability to predict traffic flow based on GPS mobility trace data in two major cities. We make use of two sets of GPS mobility trace data that were gathered from more than 500 taxis in the San Francisco Bay area for 24 days and more than 4,000 taxis in Shanghai city for 28 days. We propose an effective centrality-based methodology combined with minimal locale information in an attempt to predict the traffic flow in the road networks. We discuss four centrality metrics---namely, geodesic shortest-path betweenness, Euclidean shortest-path between, fastest-path betweenness, node-weighted fastest-path betweenness, and their relationships to traffic flow. We argue that based solely on topological properties of urban networks, as supported by previous work, provides a limited prediction power of traffic flow in a city. The accuracy can be enhanced given minimal information on location liveliness, for example by incorporating commercial and road information.

We review existing work on weighted urban network analysis, network-based traffic prediction, human mobility traces and some of their applications in real-world problems in the next section. We will then discuss a novel framework which allows node weight and travel speed to be incorporated into traditional betweenness analysis in Section \ref{5meth}; we demonstrate that it significantly improves traffic prediction performance in Section \ref{5results}. Based on the framework, we identify interesting behaviours topologically which differs in rush hour and non-rush hour traffic. We conclude with evaluation and future work in Section \ref{5conclusions}.

To the best of our knowledge, this is the first work on network-based traffic flow analysis which is based on such large-scale vehicle mobility traces in intra-city networks.


\section{Related Work}\label{5related}
Typical topology and patterns in spatial networks~\cite{Gastner2006} such as the roads, Internet, and flight networks are unlike their well studied non-spatial counterparts (e.g. social, biological, and technological networks), which disregard the physical geography of the nodes and exhibit various important properties~\cite{big}. Urban spatial networks, due to the geographical constraints, are typically planar, i.e., they can be drawn on a 2D plane without any edges crossing due presumably to construction constraints. Planar graphs can be shown to have an average degree strictly less than 6 and a diameter which grows faster  than non-spatial real-world networks ($\sqrt{N}$ as opposed to $\log{N}$ in a network of size $N$) without a well defined low dimension (See~\cite{Gastner2006}).

L\"{a}mmer et al.~\cite{Lammer2006} reported the effective dimensions ranged from 2 to 2.5 in 20 Germany city road networks as well as a power-law betweenness centrality distribution over the nodes. A deduction was then made that majority traffic volume is concentrated on a minority of roads but this was not supported by real evidence. Cardillo et al.~\cite{Cardillo2006} analysed unit-square mile tile samples of 20 different cities and reported that cities exhibited \emph{meshedness}, an alternative to the \emph{clustering coefficient} used in non-planar graphs, given by the number of faces associated with the planar graph with $N$ nodes and $M$ edges over the maximum possible. Road networks were also reported to be globally \emph{efficient}~\cite{Cardillo2006}, which meant that the actual distance required to travel from any two points in the city does not deviate too much from their straight line Euclidean distance. Attempts to characterise and compare different cities were done using these local and global properties specifically designed for spatial graphs.

Researchers have studied human mobility for the understanding of the nature of human movement which is vital to aspects pertaining to epidemic spreading and communication network design~\cite{colizza-2008-251, gonzaleznature08, huihumanmobility}. For instance, analysis of human mobility traces has demonstrated power-law inter-contact time distributions with cut off~\cite{psn-tmc07,Karagiannis07mobile}, levy-flight patterns consisting of lots of small moves followed by long jumps~\cite{Rhee2008}, etc. Vehicular mobility has also been known to be important for resource allocation and communication network optimization in a city~\cite{Bai2009}. Krings et al.~\cite{Krings:2009:SBC:1632710.1633616} used the anonymous communication patterns of 2.5 million customers of a Belgian mobile phone operator to construct a social network of customers and obtain a network of cities by grouping customers billing address (571 towns and cities). They showed that inter-city communications can be characterised by a gravity model and the intensity of communication between two cities is proportional to the products of the two populations divided by the square of the distance between the cities. Calabrese et al.~\cite{CalabreseRatti2010} experimented with platform by Telekom Italia to monitor real-time urban dynamics based on instantaneous positioning of buses and taxis.

Traffic flow analysis has indeed been subject to years of investigation from various fields. We do not discuss mathematical or agent-based computational modelling of traffic flow here, to which readers are referred to~\cite{Garavello2006}. Crucitti et al.~\cite{Crucitti2006} discussed and analysed various centrality distributions in different cities. Both the primal spatial graph and its dual topological graph were discussed. Correlation has also been established between commercial activities and node centralities~\cite{porta2009street}.

Several studies spanning from inter-city to intra-city scale have been performed. Kurant and Thiran~\cite{Kurant2006} proposed a multi-layer network framework to study and compare weighted traffic networks and the underlying physically topology. But their traffic analysis was only based on timetables of regular public transportation systems in a city. Betweenness and a modification with restrictions on the sources of nodes were used as traffic predictors. The authors reported that both predictions were insufficient to approximate node load because real life traffic pattern was highly heterogeneous.

Jiang and Liu~\cite{Jiang2009} attempted to predict traffic flow using predictors from space syntax~\cite{Ratti2004}, i.e., the node degree and \emph{integration} (equivalent to closeness with restriction of number of hops from the node). Traffic data was collected from official traffic census using inductive loop and pneumatic tube counters but are restricted to only hundreds of roads. Correlation as high as 0.8 were achieved but only when small sample areas from the entire map were selected. Attempts were also made in~\cite{Jiang2008, Jiang2009a} to correlate Google's PageRank~\cite{Page1999} on junctions to predict the respective traffic flow. Their results indicated that PageRank offered minimal advantage over simple measures based on degree or closeness. PageRank assumes a random walk model over a network which is intended to model human web browsing behaviour. However, we believe that, despite incomplete information on the road network structure, human tend to know the destination before setting off and hence the two processes are fundamentally different.

De Montis et al.~\cite{DeMontis2007} studied a weighted inter-municipal commuting network in the Sardinia region, Italy. Official statistics of traffic between major cities in the region was incorporated into the networks. High correlation between the connectivity of the municipality and commuter traffic was found.

Kazerani and Winter ~\cite{Kazerani2009} argued that shortest-path betweenness centrality was against human way-finding behaviour as most people find their way based on incomplete and inaccurate network knowledge. It was also argued that sources and sinks of traffic were irregularly distributed in the networks, a fact that is not captured by betweenness centrality which measures shortest paths between all pairs of nodes in the network. A modification to betweenness was proposed, which assumed that the exact travel demand, starting and ending points were known. This information would be extremely hard to obtain and their approach was unsubstantiated by real data.

\section{Methodology}\label{5meth}

\subsection{Network Construction}\label{5metha}

Since the San Francisco taxi traces were taken in 2008~\cite{Piorkowski2009}, we have obtained the 2008 US Government Census road shape files of the San Francisco Bay area~\cite{tiger}. The shape file contains information of every road in the city as polylines defined by their corresponding GPS coordinates. Where the GPS coordinates of any segment of two different polyline match, the coordinate is interpreted as a road intersection between the two roads. Each road is also conveniently classified by one of the ten possible road feature classes according to the US Bureau of the Census MAF/TIGER Feature Classification Code (MTFCC), of which we only keep three classes: type S1100---Interstate, type S1200---Major Highways and type S1400---Local Neighbourhood Road, Rural Road, City Street. The road network contains 26,049 nodes and 33,079 edges, spanning a total estimated distance of 1,980km.

For the analysis in Shanghai, we have obtained the network shapefiles from OpenStreetMap\footnote{www.openstreetmap.org}. OpenstreetMap is a public mapping service which allows users to update and edit the map freely. We also note with caution that the discrepancies between the map snapshot of Shanghai taken at the time of writing and in 2007 when the traces were taken are likely to introduce errors in terms of shortest-path calculations as well as potential invalid snapping of GPS traces to the roads. The road network of Shanghai contains 54,151 nodes and 61,834 edges, spanning a total estimated distance of 7,402km.

\begin{figure}[ht]
\centering
     \includegraphics[width=\columnwidth]{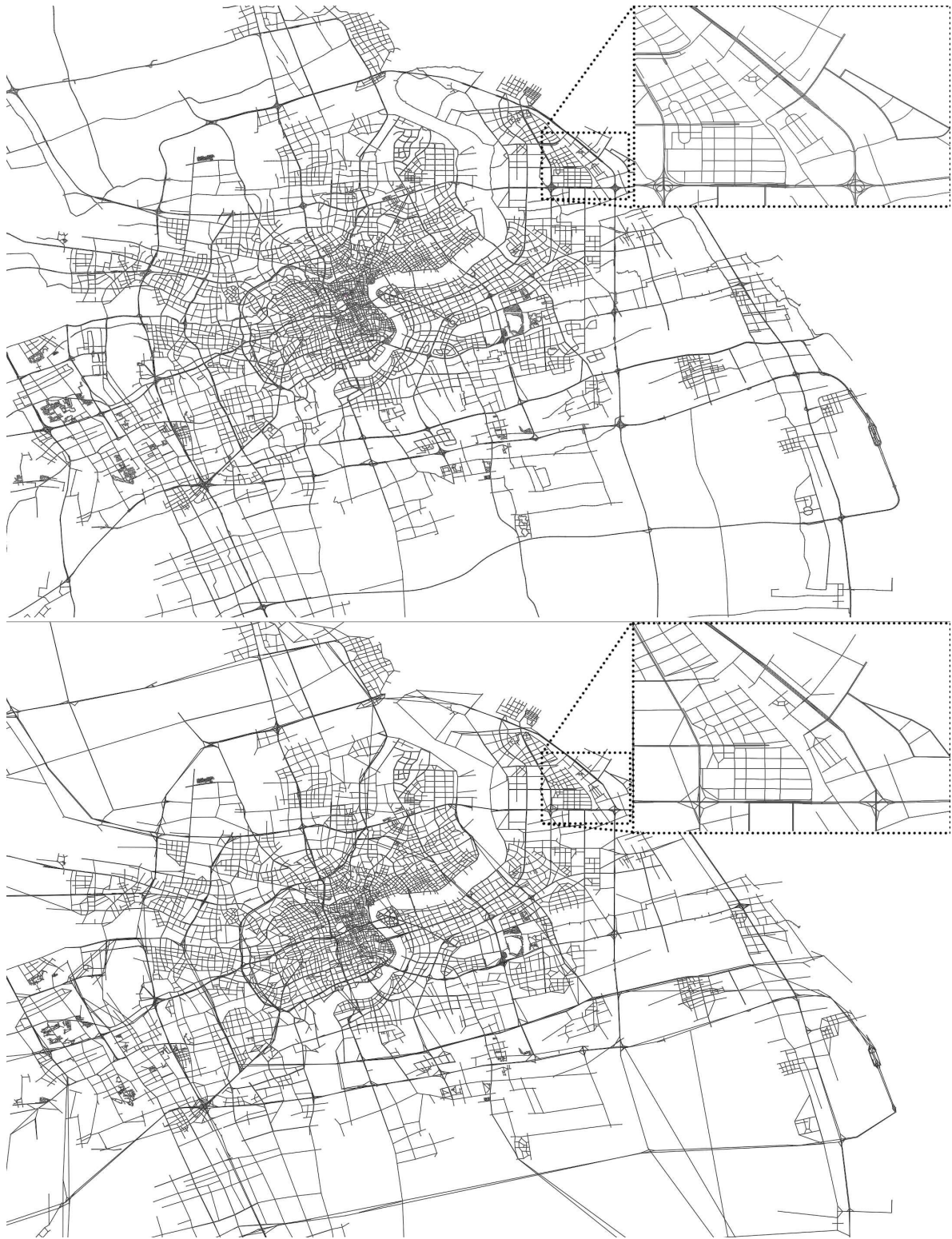}
     \caption{The original network representation of the Shanghai road network(\emph{top}) and the simplified network representation by removing all junctions with node degree 2 (\emph{bottom}).}\label{shanghai}
\end{figure}

We only take the largest connected component of the resulting networks. The networks are undirected due to data constraint. While knowing the exact flow restrictions on or between each individual road would be beneficial to our analysis, we leave it as a potential and important future work. For our analysis, we reduce the complexity and size of the network while keeping the maximum information using the following scheme. A node/junction of degree two in the original network is either the same road or a corner of two roads, which serves no other purpose than maintaining the road's correct shape on the map. Hence, for every such node B, we remove the node and the two corresponding edges (A,B) and (B,C), and establish a new link between A and C. We assign a weight to (A,C) equal to the sum of the weights of edges (A,B) and (B,C). If a link already exists between A and C, we assign the new weight to be the minimum of the old and new weights. We simplify the network for the following reasons:
\begin{enumerate}
 \item It greatly speeds up the centrality analysis.
 \item It does not alter any of the shortest-path analysis as the path lengths are maintained as described above. The key structure of the whole network is also maintained.
 \item We are interested in the traffic flow through each intersection---nodes with degree 2 are highly prevalent in the original network as they exist to maintain the correct shape of roads on the map but are irrelevant to our study.
\end{enumerate}
A graphical comparison between the original and the simplified Shanghai road networks is shown in Figure \ref{shanghai}. As we can see, the key differences between the simplified and original graph are the straightening of certain curved roads as well as mergers of certain roads which connect two identical intersections. Since our analysis will be focused on actual intersections and the number of nodes in the network is significantly reduced, we believe our road simplification technique is a valid simplification. The simplified networks for San Francisco and Shanghai contain 9,791 nodes, 16,129 edges and 12,979 nodes and 20,585 edges, respectively. Key statistics of the road networks are shown in Table \ref{5nettab}.

\begin{table}[htbp]
 \centering
\begin{tabular}{|c|c|c|c|c|c|c|c|}
 \hline
  Network & $N$ & $K$ & $\langle l \rangle$ & $W$ & $d$ & $\langle k \rangle$ \\ \hline
  San Francisco & 26,049 & 33,079 & 59.9m & 1,980km & 6.67km & 2.54  \\
  San Francisco (Simplified) & 9,791 & 16,129 & 116.4m & 1,877km & 6.60km & 3.29  \\
  Shanghai & 54,151 & 61,834 & 119.7m & 7,402km & 20.87km & 2.28 \\
  Shanghai (Simplified) & 12,979 & 20,585 & 354.1m & 7,239km & 18.9km & 3.17  \\
 \hline
\end{tabular}
\caption{Key statistics of the two road networks studied. All centrality-based analyses are carried on the simplified version of the networks as described in Section \ref{5metha}. Here, $N$ denotes the number of nodes, $K$ is the number of edges, $\langle l \rangle$ is the average edge length, $W$ is the total edge lengths, $d$ is the characteristic path length (average length of all shortest paths) and $\langle k \rangle$ is the average degree.}
\label{5nettab}
\end{table}

\subsection{Traffic Flow Estimation}

The trace for San Francisco consists of roughly 10 million GPS readings (latitude, longitude) from more than 500 taxis in the Bay area over a period of 24 days from May 17, 2008 to June 10, 2008~\cite{Piorkowski2009}. The trace for Shanghai contains more than 100 million GPS readings from more than 4 000 taxis over the entire month of February, 2007~\cite{Huang2007}.

We first remove all the traces which were recorded when the taxis were unoccupied. We believe this provides a fairer measure of aggregated traffic flow given by the actual need of passenger travels. It is believed that unoccupied taxis are highly incentive-oriented and tend to remain in certain areas where passengers are easier to be found (such as commercial areas, stations, airports, etc). Hence, including those traces in the traffic flow estimate is likely to introduce bias. Secondly, by removing traces when taxis are unoccupied we also remove meaningless data, e.g. when they are waiting at the taxi stops.

It is our assumption that these large number of taxi traces aggregated over such a long period of time provides a fair representation of road traffic flow in the intra-city urban network. However, besides the lack of samples, traffic flows generated by taxis may contain a bias on destinations where people tend to use taxis (such as tourist and major business spots). Moreover, the route choice of taxi drivers may not necessarily reflect those chosen by typical drivers; it is likely also that commuter traffic, which contributes greatly to the city's overall traffic, may not be captured by taxis. Nonetheless, given a limited amount of resources, and as opposed to the extreme end of collecting GPS traces of every vehicle in the city, we believe that taxis are the more representative class of vehicles for measuring intra-city traffic than private vehicles or public transport systems with pre-defined routes. We remain cautious that traffic flow generated by a sample of taxis may fail to capture the entirety of traffic flow in a city.

To effectively estimate the actual traffic flow (the number of unique cars that goes through a junction given a period of time), it is important to distinguish the paths that are taken by each unique taxi at different times. In both cities, the average interval between each pair of consecutive traces is roughly 1 minute and the average speed is below 30km/h.  We first define the notion of a \emph{trip} to be a sequence of GPS traces of a particular taxi in which each consecutive trace pair is less than 90 seconds apart and implies a speed $<= 120km/h$. Then, for every such valid trace in each trip, we locate its closest edge and increase the traffic count of the two corresponding junctions by one. By definition, each unique trip can only visit a junction once and hence traffic count for each unique junction can only be increased once per trip.

Due to the time resolution of the data, it is often the case that each consecutive trace pair is several junctions apart. If we follow the simple scheme above, we risk losing valuable information of the entire trajectory taken by each taxi during the trip. We employ two methods to interpolate the trajectory of the traces when they are too far apart.

\subsubsection{Fastest-Path Interpolation}
We first attempt to estimate the actual edge traversed based on the fastest path between the two nodes. Where two nodes are more than one hop apart, we run Dijkstra's algorithm from either end to devise a fastest path between them. The final path profile is a sequence of connected junctions which approximates the actual path of the taxi during that trip. Similarly, we increase the traffic flow count of each junction in the interpolated path by one. This estimation technique is based on the assumption that taxis always take the shortest/fastest route from source to destination. Unfortunately, this interpolation is biased towards our betweenness-based prediction (Section \ref{5fast}) as betweenness is itself based on the shortest/fastest-path assumption. Since the average speed and time interval indicate an average distance of 500m between each consecutive trace, this estimation is highly vulnerable to the bias especially in the case where the trajectory is not straight.

\subsubsection{Interpolation by Routing Service}
Our second technique involves interpolating the trajectory of the entire trip by submitting each consecutive trace pair to the MapQuest Directions API service\footnote{http://developer.mapquest.com/web/products/open/directions-service} with the default ``Fastest Route" option enabled. Given the comprehensive information on each individual road which online routing services such as MapQuest possess, we believe that its fastest-path predictions can keep the errors to a minimal. Indeed, it is implicitly assumed that the route returned is the one picked by the taxi driver which again is not always true. However, we believe that the resolution of the traces and the criteria we employ to merge the traces into individual trips keep the potential errors to a minimum while retaining as much traffic information as possible.

Here, the interpolated trajectory for each trip is subjected to further filtering as it is noted that some routes returned by MapQuest were affected by GPS snapping errors. In some cases, a route was suggested which would mean the taxi had travelled the entire a length of the road and to come back to a point nearby, just because the consecutive GPS trace was snapped to the same road going in the opposite direction. Again, a threshold is set such that the interpolated trip could not imply a travel speed of more than 120km/h by the taxi.

\begin{figure}[!htb]
\centering
  \includegraphics[width=0.8\columnwidth]{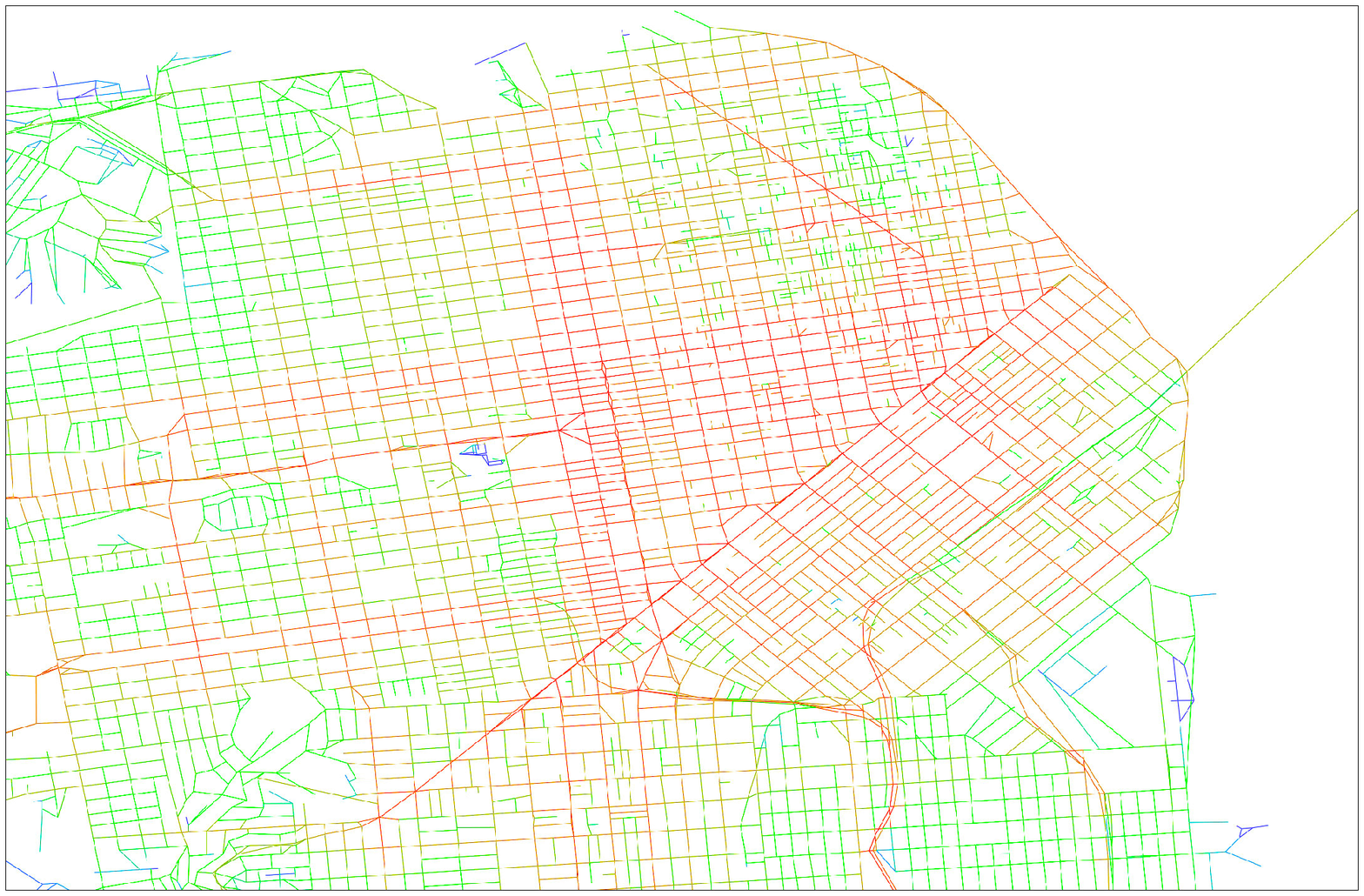}\\
  \includegraphics[width=0.8\columnwidth]{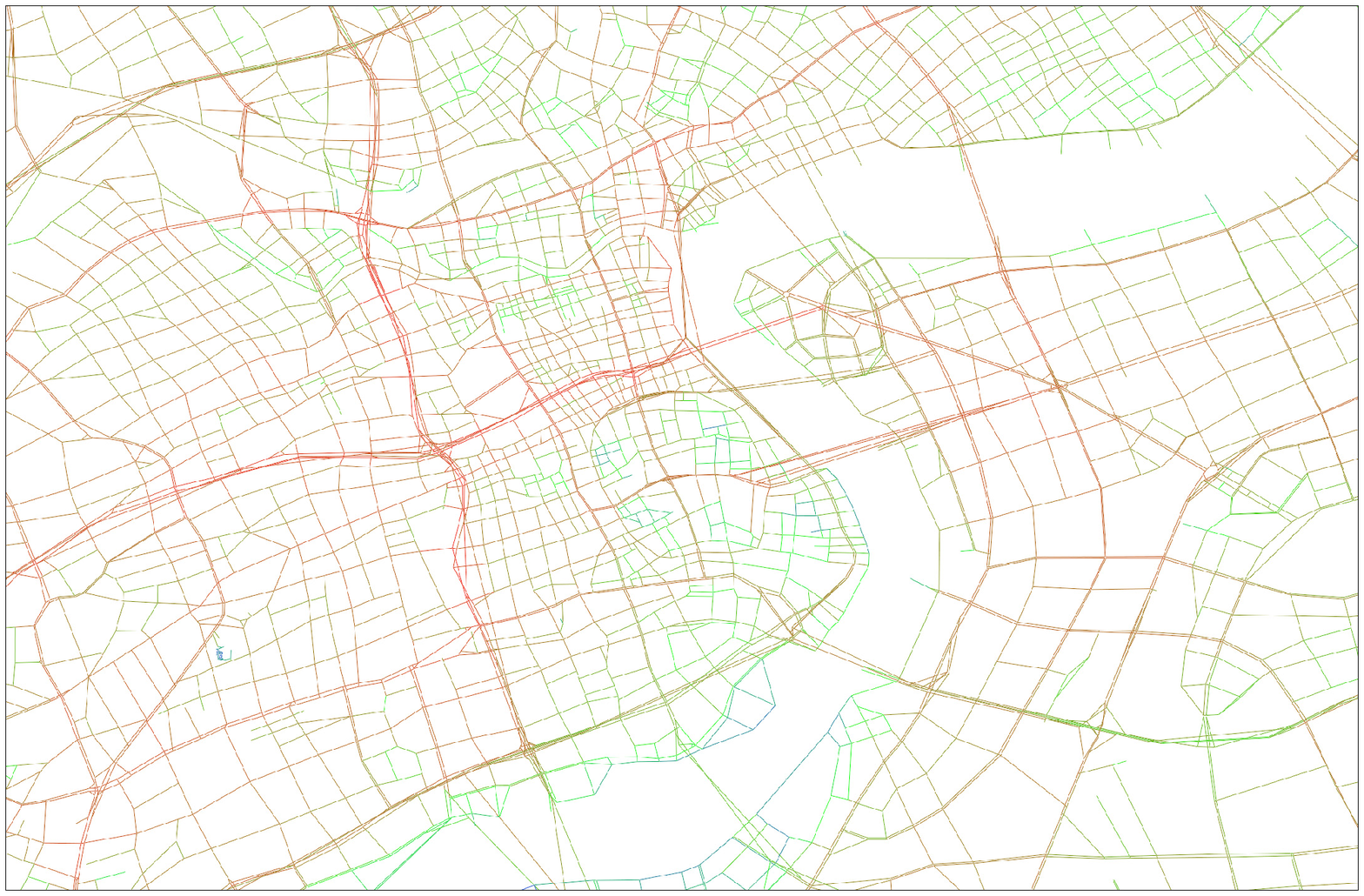}
    \caption{Traffic flow (interpolated by MapQuest routing service) in the commercial district of San Francisco (\emph{top}) and Shanghai (\emph{bottom}). For visualisation purpose, the traffic count for each edge is given by the average of its two respective ends. Red coloured edges correspond to a high number of traffic counts and blue colour corresponds to a close to zero traffic count. }\label{5traffic}
\end{figure}

Finally, for each valid trip, the traffic count on each of the distinct interpolated junctions is increased by one. This is similarly based on the assumption that in a single trip, a taxi does not visit the same junction twice. Figure \ref{5traffic} depicts graphically the overall traffic flow in the key commercial districts of the two cities based on this interpolation technique.

\begin{figure*}[!htb]
\centering
  \includegraphics[width=0.8\columnwidth]{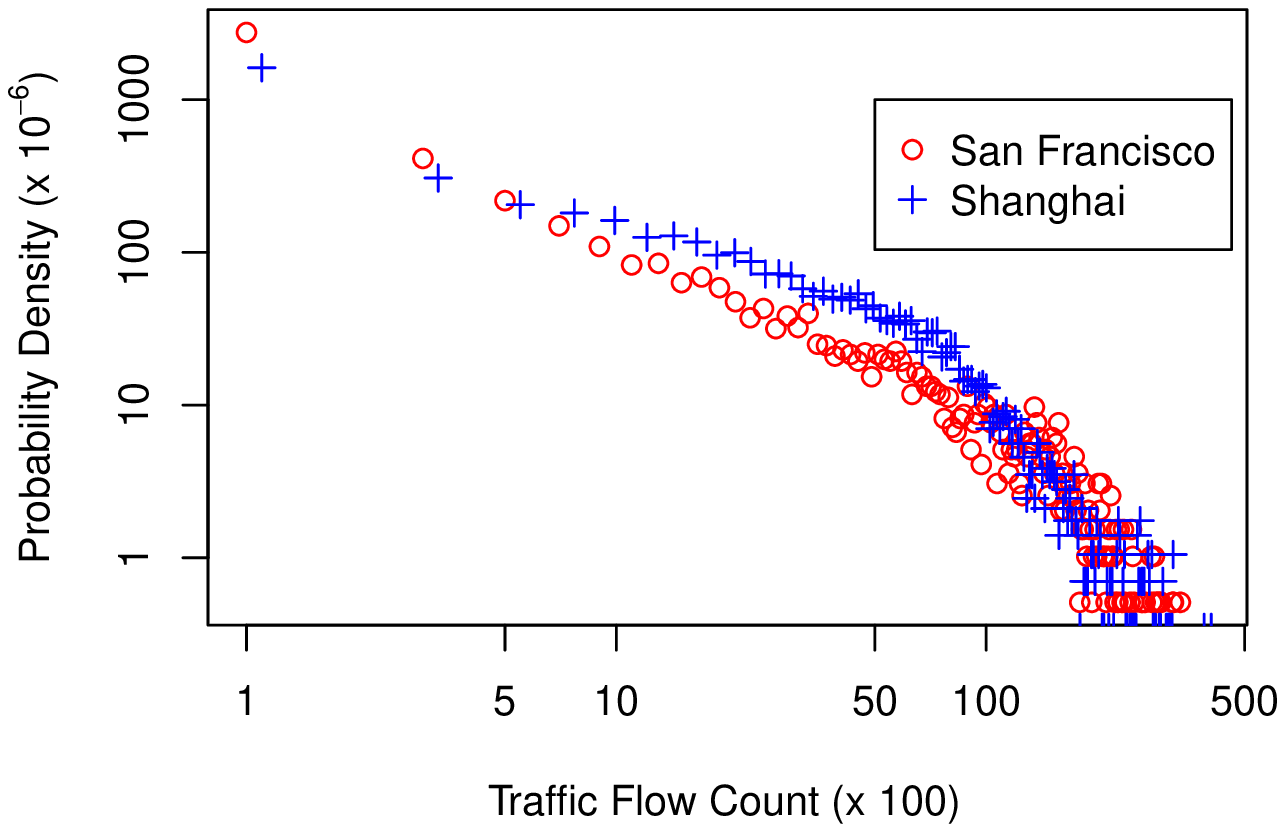}
  \caption{Traffic flow distribution per junction in San Francisco and Shanghai, with trajectories interpolated by MapQuest routing service. }\label{5trafficflow}
\end{figure*}

Figure \ref{5trafficflow} shows a distribution of traffic flow count per junction in the two cities. A power-law decay trend with a cut off (which can be explained by the finite nature of traffic flow count) is observed. Also, it indicates that only a tiny fraction of junctions carry a relatively substantial traffic load in both cities.

\subsection{Centrality Analysis}
Both degree and closeness centralities were extensively used under the study of Space Syntax, and in many cases were found to significantly correlate to traffic flow, safety against criminality, commerce activity, activity separation and pollution~\cite{Crucitti2006}. We shall present the experimental results including the aforementioned centrality measures in the next section. Here, we focus our discussion on betweenness centrality as well as some if its extensions for traffic networks which will be introduced in due course.

\subsubsection{Geodesic shortest-path betweenness}
\emph{Geodesic shortest-path betweenness} refers to the simplest definition of betweenness, i.e., the fractions of shortest paths between any pairs of nodes in the network which go through a particular node:
\begin{equation}\label{2bet}
  C_{B}(v) = \sum_{u,t \in V}{\frac{\sigma_{ut}(v)}{\sigma_{ut}}},
\end{equation}
where $V$ is the set of nodes in the network, $\sigma_{ut}$ is the number of shortest paths between nodes $u$ and $t$, and $\sigma_{ut}(v)$ is the number of those going through node $v$. Here the network is treated as a non-spatial network, where each edge corresponds to a single hop from one node to another with equal weight. For its derivation, we employ the original Brandes algorithm~\cite{brandes2001} which carries out a breath-first search from each node in the network and accumulate the shortest path counts on each node that is on the shortest path.

\subsubsection{Euclidean shortest-/fastest-path betweenness}\label{5fast}
\begin{figure}[!htb]
\centering
  \includegraphics[width=0.8\columnwidth]{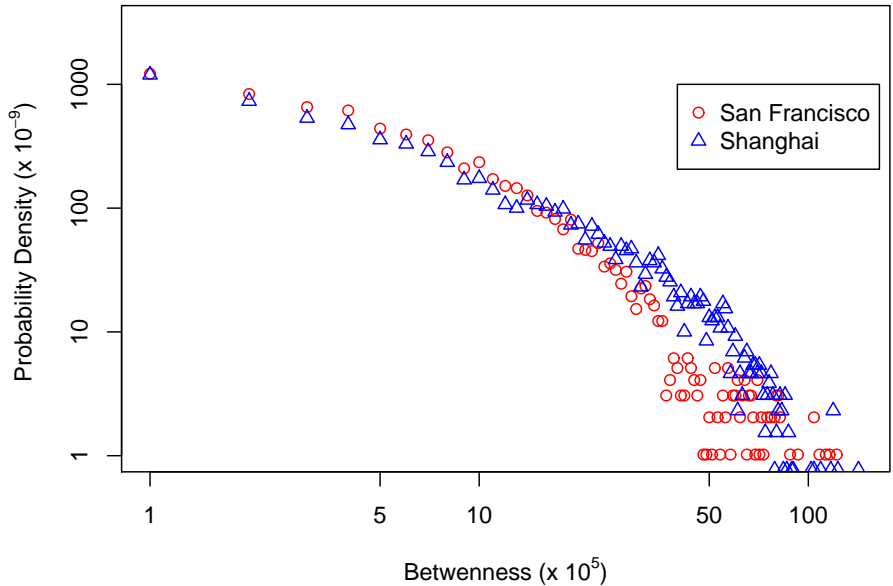}
  \caption{Distributions of Euclidean shortest-path betweenness centrality in the two cities.}\label{5dist}
\end{figure}

The drawback of geodesic shortest-path is that it does not take into account the physical distance of the spatial network when devising the shortest path. It is highly sensitive to short edges which are common in spatial networks due to the large number of intersections of different roads. Traversing on such networks generally requires a relatively large number of hops compared to non-planar networks. A simple way to circumvent this problem is to take into account the Euclidean distance of each edge when calculating all the possible pairs of shortest paths. We denote this as the \emph{Euclidean shortest-path betweenness}. The distributions of this betweenness of each junction in the two cities are shown in Figure \ref{5dist}.

One can further improve the resemblance to traffic decision by including speed consideration of different roads in the city. As with most route decision made by human or computer, the fastest route is often favoured over the shortest route. We therefore experiment with replacing the weight of each edge by an estimate of the travel time required based on its pre-labelled feature class. In most modern cities, the speed limit in typical residential or city streets is 50km/h while that in highways normally doubles the former. For the sake of brevity, we halve the weights of all edges specified as interstate/highways/primary/trunk roads in the network to accommodate for an average driver decision on choosing the faster routes. Understandably, route decision should also depend greatly on the time of day, the width, permeability~\cite{Ratti2004}, or potential charges on using the routes. A full analysis would therefore require a better understanding of route decision made by every individual. We call this \emph{Euclidean fastest-path betweenness}. Similar attempts in capturing betweenness which the shortest-path assumption does not always hold have been investigated in the realms of packet routing in communication networks~\cite{Dolev2010}.

For both the aforementioned measures, we employ the weighted version of Brandes algorithm for its derivation, with all nodes set to the same weight (see Algorithm \ref{5algor1}).

\subsubsection{Node-weighted fastest-path betweenness}
It has soon come to our attention that pure topological measures which ignore locale information would fail to capture the actual traffic flow due simply to the fact that, after all, places or nodes which are more densely populated or commercially active inevitably generate and attract more traffic. Without resorting to a full blown population or commercial density map and in the spirit of using as little accessible information as possible, we decide to choose restaurant density as a measure to give each node a corresponding weight. Restaurants and their exact locations are readily available information on a lot of online or offline directories. While there is a potential bias to restaurants which publish themselves online, we believe it provides a fair estimate of traffic generation and attraction as restaurants are designed to serve in places where people are around.

For each restaurant, we pre-assign to each junction within its 150m radius a weight inversely proportional to the number of junctions in that radius. This assumes that every restaurant carries the same traffic attracting and generating factor which is spread evenly across junctions within its 150m radius. According to~\cite{porta2009street}, 100m, 200m, and 300m are common choices of geographical granularity in the literature which respectively correspond to the scale of a street, block, and neighbourhood. Importantly, we stress that we do not by any means try to imply that the density of restaurants should be used as the node weight. Rather, we hope to emphasize the importance to capture the aspect of nodes generating/attracting different amount of traffic, and that one of the potential measure can be something as simple as the number of restaurants.

The \emph{node-weighted fastest-path betweenness} framework follows closely the algorithm proposed in~\cite{Chan2009} which is based on a modification of the original Brandes algorithm to take into account of node weights. We present the full algorithm in Appendix A. The key observation is on \textbf{line \ref{5lineA}} of the algorithm, where we adjust the number of shortest-path counts based on the node weights during the back propagation step from destination to source. For the sake of brevity, our assumption is that the overall traffic between two nodes can be estimated by the multiple of the two respective weights.

\begin{figure}[ht]
\centering
     \includegraphics[width=\columnwidth]{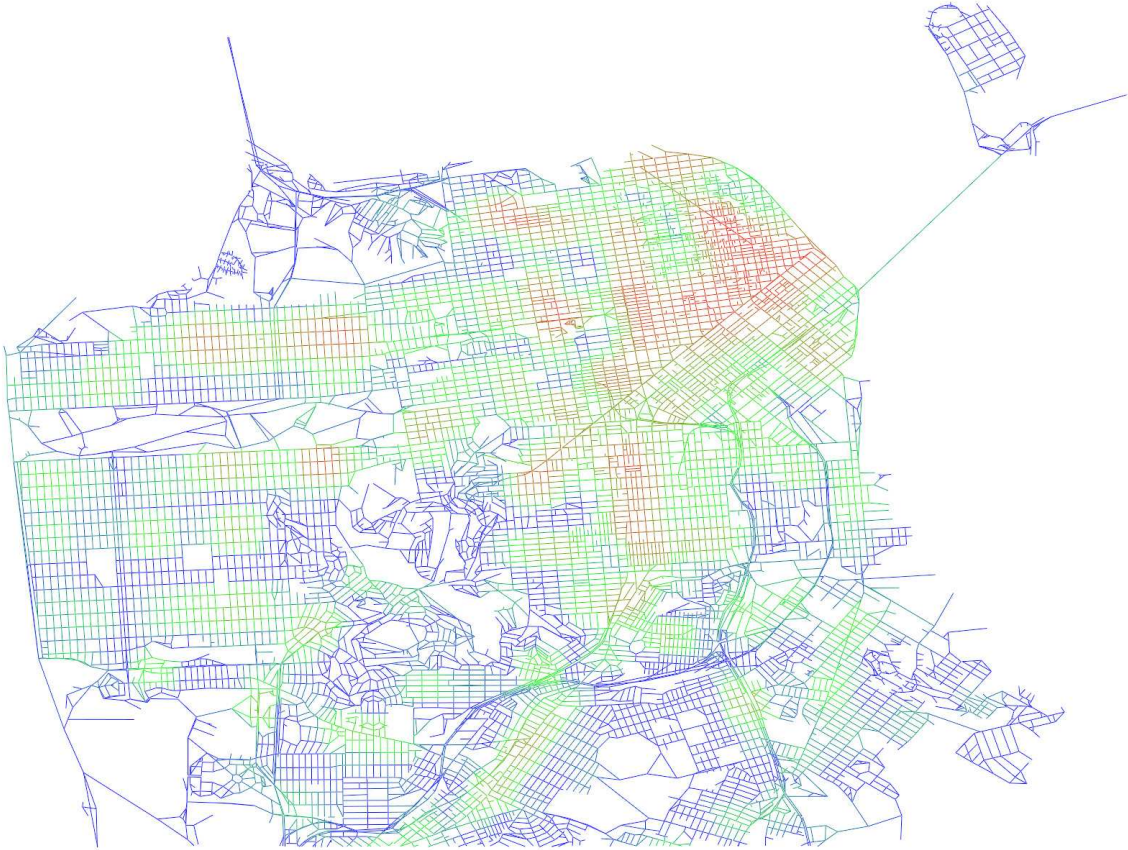}
     \caption{Density of restaurants in the San Francisco Bay area. For visualisation purpose, the restaurant count for each edge is given by the average of the number of restaurants surrounding a 150m radius from its two respective ends. Red coloured edges correspond to a high number of restaurants in its surroundings and blue colour corresponds to insignificant number of restaurants.}\label{restaurants}
\end{figure}

As an example, we have located 4,066 and 53,748 restaurants that are within the area of interest in San Francisco and Shanghai respectively. Figure \ref{restaurants} shows the density of restaurants in the San Francisco Bay area. Note that again these restaurants are current information and therefore may not reflect truly the distribution at the time the traces were taken.

\subsection{Correlation Analysis}
We evaluate the quality of predictions by devising the Pearson correlation coefficient between the traffic count and the predicted magnitude. The Pearson correlation coefficient, or Pearson's \emph{r}, is a value ranged from -1.0 to 1.0 which measures the linear dependence of two variables. A value close to zero means that the two variables are not correlated at all, i.e., knowing one does not tell much about the other. Conversely, a value close to -1.0 or 1.0 indicates that there is a perfect linear relationship between the two variables. For each junction, we therefore calculate 7 different measures for comparisons: its degree, closeness\footnote{The notion of closeness used in our analysis is different to space syntax's global integration in that distance is calculated based on the fastest-path weight between nodes.}, number of restaurants in its 150m radius, and the four betweenness measures.

While node-based correlation is simple and intuitive, it has several drawbacks. Firstly, it is susceptible to noise caused by routing inaccuracy due to insufficient information of the road structure. Consider a multi-lane highway which would typically be depicted as a number of parallel edges in the network. Disregarding the directional or turning restrictions, the shortest-path based estimation would often pick one and only one of these parallel edges to increase the traffic count. Also, routing errors mentioned in the last section due to insufficient time resolution of the GPS traces are bound to lead to noise and inaccuracies in node-based correlations. Finally, despite having removed all junctions of degree 2 from the network, certain remaining junctions which are physically close to each other in the network might exhibit similar behaviours in terms of traffic flow as well as centrality. Therefore, blindly carrying out node-based correlations may be biased to nodes which are located closely.

To improve on these issues, we follow a similar scheme used in~\cite{porta2009street} by carrying out spatial smoothing on the predictions and traffic counts prior to the correlation, a common technique known as Kernel Density Estimation (KDE). In essence, for any point on the map, a measure is obtained by summing up all the events occurred on the whole 2D plane weighted inversely by their Euclidean distances from that point. To achieve that, a kernel function $\kappa$, whose shape is symmetric and integral is one, is added to an overall sum over the plane where an event occurs. The sum therefore forms a surface over the plane with a volume equivalent to the number of events. It is typically normalised such that it becomes a probability density function over the 2D plane. For a given point on a 1-dimensional line with coordinate $x$, the measure by KDE is commonly given as follows:
\begin{equation}\label{kde1}
 \hat{f}(x) = \frac{1}{n} \sum_{i=1}^{n}{\kappa\left(\frac{(x-x_i)}{h}\right)},
\end{equation}
where $n$ is the total number of events in the entire space, $h$ is the bandwidth, and $x_i$ is the coordinate of the event $i$. $h$ acts as the smoothing factor---the larger $h$ is, the more likely a distant event has an effect on the point concerned.

Since the coordinates are real-valued, it is common to discretise the 2D plane into pixels or squares prior to running KDE. We follow closely the settings used in~\cite{porta2009street}, setting each pixel to be $10m$ by $10m$, and $h = 300m$. We employ the standard Gaussian kernel given by:
\begin{equation}\label{kde2}
 \kappa(x) = \frac{1}{\sqrt{2\pi}}\, e^{- \frac{\scriptscriptstyle 1}{\scriptscriptstyle 2} x^2}.
\end{equation}

To ease the calculation, we pre-calculate the values of the six predictors, traffic count, as well as restaurant count for each 10m by 10m pixel. Where there is more than one node in the pixel, we take the sum of the node-based measures as an aggregated estimate for the pixel. For restaurant count, the number of restaurants within the pixel is used instead. To further speed up the process, pixels which are beyond the bandwidth $h$ from the pixel concerned are not considered in the summation.

To summarise, we first pre-calculate the aggregated estimates for every pixel on the map. Then, for each pixel with coordinates $\textbf{x}$ which contains any junction, we devise the following metric:
\begin{equation}\label{kde3}
 \tilde{f}(\textbf{x}) = \sum_{\{\textbf{x}_i| d(\textbf{x},\textbf{x}_i) \leq \frac{h}{2}\}}{w_i \kappa\left(\frac{d(\textbf{x},\textbf{x}_i)}{h}\right)},
\end{equation}
where $ d(\textbf{x},\textbf{x}_i)$ is the Euclidean distance between the concerned pixel with coordinate $\textbf{x}$ and another pixel $i$ with coordinate $\textbf{x}_i$; $w_i$ is the pre-calculated value (e.g. the aggregated centrality or traffic count) of pixel $i$.

The correlation analysis only takes into account pixels which contain junctions. We do not correlate pixels that contain only edges (roads) since our knowledge of the edges, i.e., their traffic flows and centralities, is no more than that of their corresponding junction ends. Hence, given the framework, correlating every pixel which contains an edge may yield an overall biased Pearson correlation.

Note that the above measure transforms the effect of an event on the 2D space using a 1-dimensional Gaussian as we assume an event's effect is symmetric in any direction on the 2D plane. We do not normalise $\tilde{f}(\textbf{x})$ as it does not affect the correlations.

\section{Results and Discussion}\label{5results}
We present the results for our correlation analyses of San Francisco and Shanghai in Tables \ref{5tab1} and \ref{5tab2} respectively. For each entry, both the node-based and KDE pixel-based Pearson's $r$ are included.

\begin{table}[htbp]
 \centering
   \begin{tabular}{l|ccc}
    \multicolumn{4}{c}{\textbf{San Francisco} - Pearson's $r$ (Node-based/KDE)} \\
    No Interpolation & Non-Rush Hour & Rush Hour & Overall \\ \hline
    Degree & 0.272/0.392 & 0.246/0.388 & 0.268/0.394 \\
    Closeness & 0.243/0.413 & 0.19/0.404 & 0.231/0.413 \\
    Geodesic betweenness & 0.236/0.189 & 0.191/0.142 & 0.226/0.177 \\
    Shortest-path & 0.273/0.333 & 0.201/0.271 & 0.256/0.318 \\
    Fastest-path & 0.264/0.315 & 0.203/0.249 & 0.249/0.298 \\
    Node-weighted fastest-path & 0.591/0.83 & 0.526/0.767 & 0.579/0.818 \\
    Restaurant & 0.445/0.717 & 0.458/0.709 & 0.453/0.72 \\
          &       &       & \\
    Fastest-Path Interpolation &       &       & \\ \hline
    Degree & 0.268/0.521 & 0.267/0.538 & 0.269/0.526 \\
    Closeness & 0.256/0.529 & 0.275/0.545 & 0.261/0.534 \\
    Geodesic betweenness & 0.204/0.266 & 0.222/0.314 & 0.208/0.277 \\
    Shortest-path & 0.252/0.417 & 0.274/0.469 & 0.257/0.429 \\
    Fastest-path & 0.386/0.456 & 0.437/0.531 & 0.398/0.473 \\
    Node-weighted fastest-path& 0.565/0.753 & 0.549/0.725 & 0.564/0.75 \\
    Restaurant & 0.419/0.622 & 0.381/0.568 & 0.412/0.613 \\
          &       &       & \\
    Routing-Service Interpolation &       &       & \\ \hline
    Degree & 0.218/0.486 & 0.231/0.46 & 0.223/0.48 \\
    Closeness & 0.239/0.501 & 0.273/0.484 & 0.25/0.498 \\
    Geodesic betweenness & 0.159/0.237 & 0.179/0.23 & 0.166/0.236 \\
    Shortest-Path & 0.224/0.392 & 0.259/0.397 & 0.235/0.395 \\
    Fastest-path & 0.223/0.383 & 0.25/0.381 & 0.232/0.383 \\
    Node-weighted fastest-path & 0.528/0.764 & 0.573/0.812 & 0.544/0.779 \\
    Restaurant & 0.465/0.65 & 0.463/0.66 & 0.467/0.654 \\
   \end{tabular}%
   \caption{Pearson product-moment correlation coefficients of 6 centrality predictors and restaurant count against traffic flow in San Francisco. Traffic flow is further separated into rush hour (6AM - 10AM and 4PM - 7PM) and non-rush hour for comparisons.}
 \label{5tab1}%
\end{table}%

\begin{table}[htbp]
 \centering
   \begin{tabular}{l|ccc}
   \multicolumn{4}{c}{\textbf{Shanghai} - Pearson's $r$ (Node-based/KDE)} \\
    No Interpolation & Non-Rush Hour & Rush Hour & Overall \\ \hline
    Degree & 0.253/0.404 & 0.273/0.413 & 0.261/0.408 \\
    Closeness & 0.339/0.466 & 0.367/0.474 & 0.35/0.47 \\
    Geodesic betweenness & 0.153/0.505 & 0.173/0.516 & 0.16/0.51 \\
    Shortest-path & 0.322/0.642 & 0.35/0.637 & 0.333/0.641 \\
    Fastest-path & 0.262/0.684 & 0.29/0.676 & 0.272/0.683 \\
    Node-weighted fastest-path & 0.289/0.634 & 0.311/0.622 & 0.297/0.631 \\
    Restaurant & 0.256/0.44 & 0.276/0.436 & 0.263/0.44 \\
          &       &       & \\
    Fastest-Path Interpolation &       &       & \\ \hline
    Degree & 0.265/0.439 & 0.269/0.438 & 0.267/0.439 \\
    Closeness & 0.418/0.501 & 0.414/0.496 & 0.418/0.5 \\
    Geodesic betweenness & 0.167/0.568 & 0.179/0.577 & 0.171/0.572 \\
    Shortest-path & 0.387/0.646 & 0.388/0.631 & 0.389/0.642 \\
    Fastest-path & 0.481/0.744 & 0.476/0.724 & 0.48/0.738 \\
    Node-weighted fastest-path & 0.491/0.674 & 0.472/0.641 & 0.486/0.664 \\
    Restaurant & 0.329/0.39 & 0.321/0.377 & 0.328/0.386 \\
          &       &       & \\
    Routing-Service Interpolation &       &       & \\ \hline
    Degree & 0.313/0.505 & 0.309/0.507 & 0.312/0.506 \\
    Closeness & 0.552/0.583 & 0.55/0.586 & 0.552/0.584 \\
    Geodesic betweenness & 0.086/0.429 & 0.089/0.425 & 0.087/0.428 \\
    Shortest-path & 0.484/0.751 & 0.478/0.746 & 0.483/0.75 \\
    Fastest-path & 0.43/0.814 & 0.427/0.807 & 0.43/0.812 \\
    Node-weighted fastest-path & 0.528/0.832 & 0.515/0.817 & 0.525/0.828 \\
    Restaurant & 0.438/0.444 & 0.433/0.438 & 0.438/0.443 \\
   \end{tabular}%
   \caption{Pearson product-moment correlation coefficients between 6 centrality predictors and restaurant count against traffic flow in Shanghai. Traffic flow is further separated into rush hour (6AM - 10AM and 4PM - 7PM) and non-rush hour for comparisons.}
 \label{5tab2}%
\end{table}%

A first glance at the results reveals that predictors work differently in the two cities and across the different interpolation techniques. This is understandable given that the two cities have fundamental differences in terms of design and planning---San Francisco has a well defined grid like structure while Shanghai road structure is more irregular and ad-hoc (self-organised). As explained earlier, the fact that fastest-path interpolation would by definition favour the fastest-path betweenness predictor is also evident across both tables. KDE-pixel based correlation has in general a much higher correlation coefficient over the node-based counterpart. As discussed, due to imperfect information on the road networks and from the raw traces, carrying out spatial smoothing on the observed traffic flow and predictions would remove some of the noise from these errors. Nonetheless, we are able to draw several important conclusions from our experiments.

Degree and geodesic betweennesses offer minimal prediction power. This is as expected as the variation of degree centrality in spatial networks is so small that its discrimination power would be low. Geodesic betweenness does not take into account edge weights and hence is inapplicable for flow prediction in spatial networks where the number of hops between source and destination carries little meaning. These coincide with numerous previous findings that simple topological based measures seem to have an inherent limit in characterising traffic flow. Closeness centrality has a marginal advantage over degree centrality in Shanghai but its prediction power remains modest in both cities.

The advantage for incorporating the notion of fastest path in betweenness is minimal in San Francisco but is more evident in the Shanghai network. We conjecture that in San Francisco the use of the primary roads are occluded by the fact that we have only focussed on the northern bay area. Fastest-path betweenness may not necessary consider the use of the highways such as the Interstate 80 and 280 even though we have assigned them a favourable weight because the southern San Mateo areas to which they lead are outside of the Bay area concerned. On the other hand, the studied area in Shanghai covers the entirety of several major ring roads and trunk roads of the city which could potentially explain the difference. We consider the effect of the span of mobility traces and size definition of an appropriate ``intra-city" urban network for such analysis as important future work.

Finally, node-weighted fastest-path betweenness significantly outperforms other predictors in the routing-service interpolated analysis in both cities. The advantage of node-weighted fastest-path is much more evident in San Francisco than in Shanghai due to the fact that restaurant density in itself is in fact a reasonable predictor of traffic flow. Combining its weight to betweenness therefore produces an even more powerful predictor. This supports our notion that combining node weights and betweenness is essential to provide a consistent and accurate representation on traffic flow in a city.

Another remarkable observation is the fact that in four out of six settings, centralities have a higher correlation during non-rush hour than rush hour. This could mean that drivers tend to avoid the busiest roads or the fastest routes during peak hours. Whether drivers tend to take the shortest path during non-rush hour would vary across cities and would be an interesting phenomenon that requires further investigations with more refined data.

On the whole, as opposed to some other studies, we believe that the shortest-path assumption with appropriate weighting of source and destinations can provide an accurate prediction traffic flow in a city. Ideally, the node weight should correspond to the traffic generating and attracting power of a specific junction/location. In our case, we have demonstrated using restaurant density as an example candidate for weighting each node. Obviously, utilising all possible locale information such as population density and commercial activity could be useful but is against the minimalist spirit of the framework. Nonetheless, in a hypothetical city planning scenario, although it may not be possible at all to predict the emergent human activity, we foresee that factors such as land use planning can be more effectively incorporated into transport planning using the node- and edge-weighted betweenness framework.

\begin{figure}[htb]
\centering
   \includegraphics[width=0.49\columnwidth]{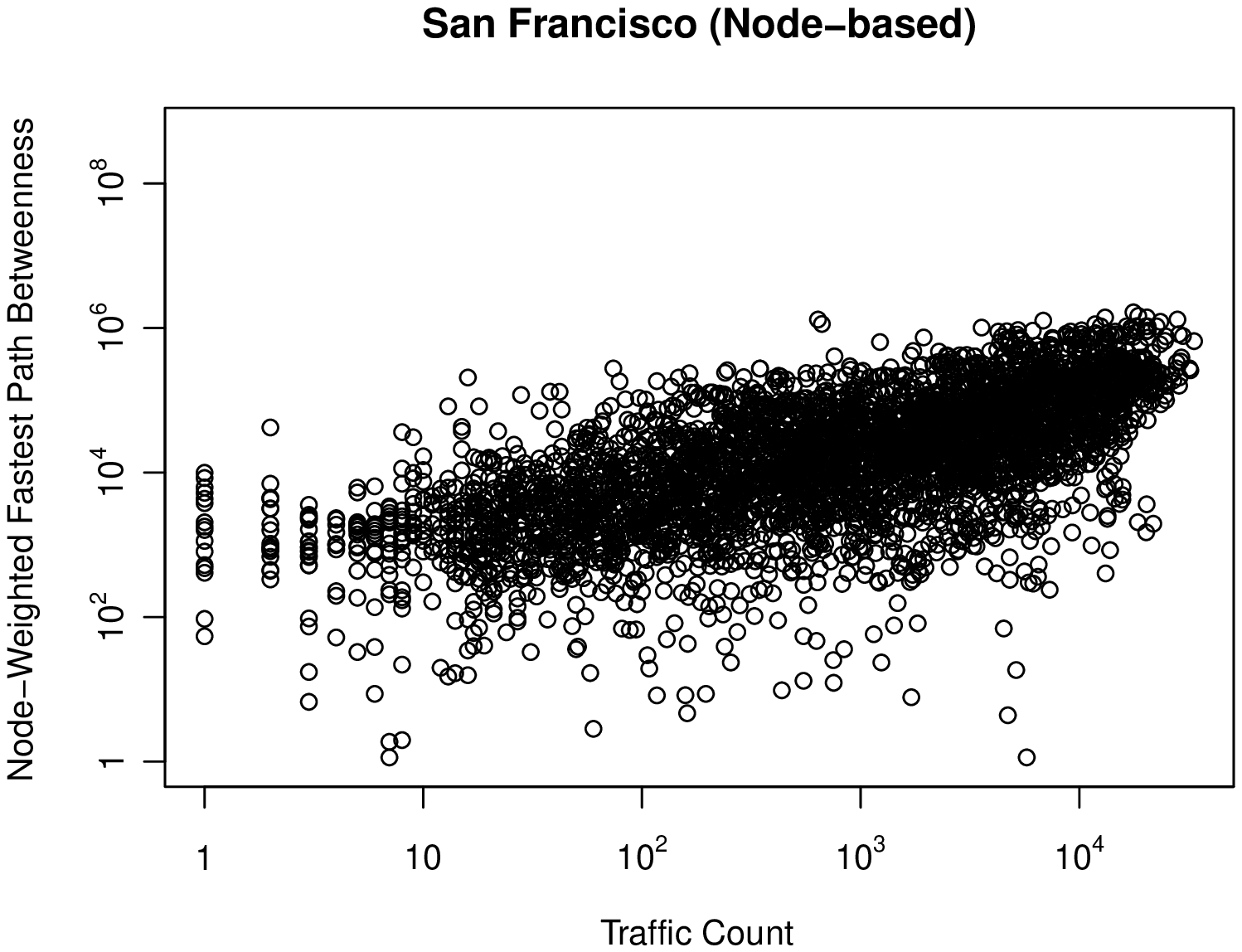}\includegraphics[width=0.49\columnwidth]{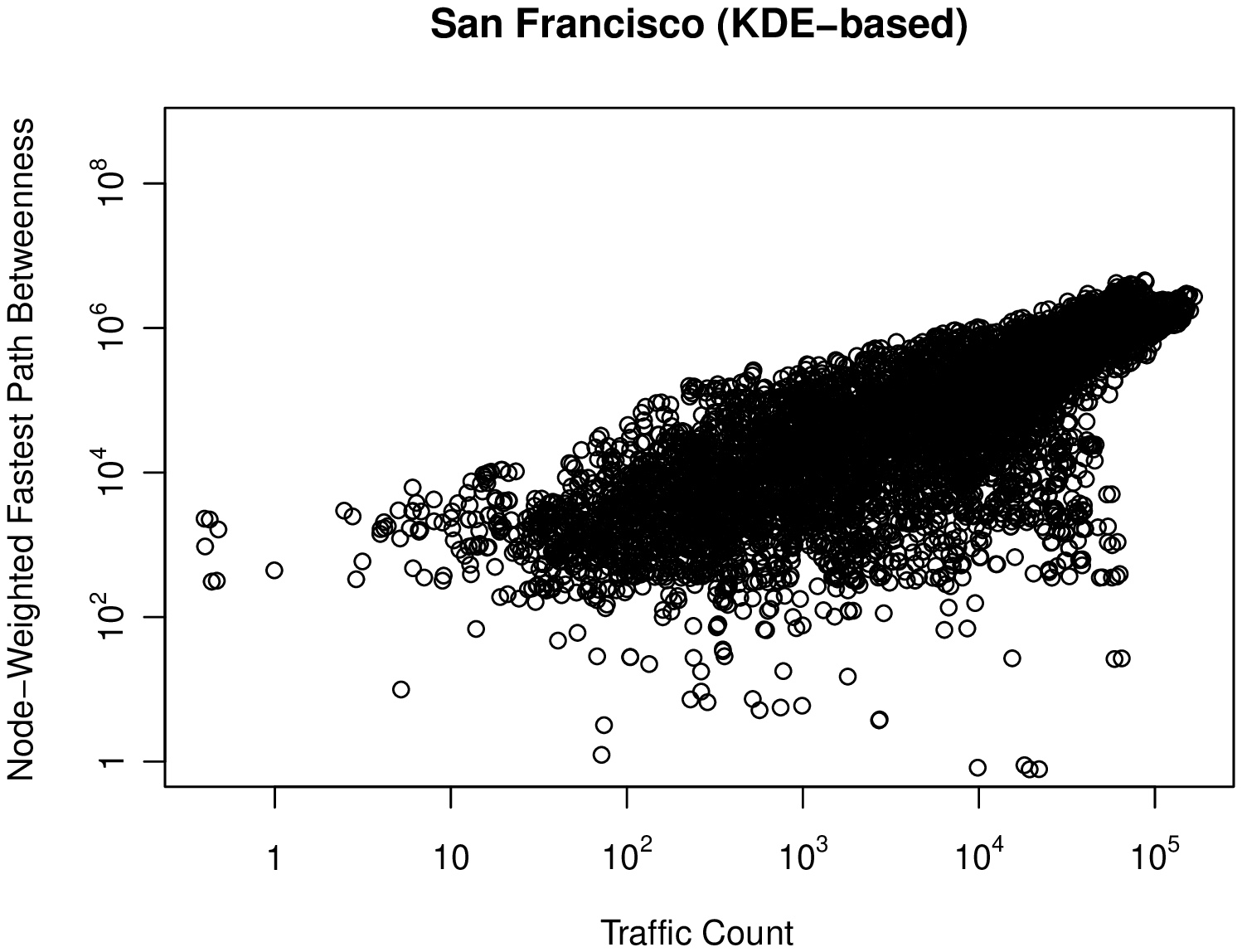}\\
   \includegraphics[width=0.49\columnwidth]{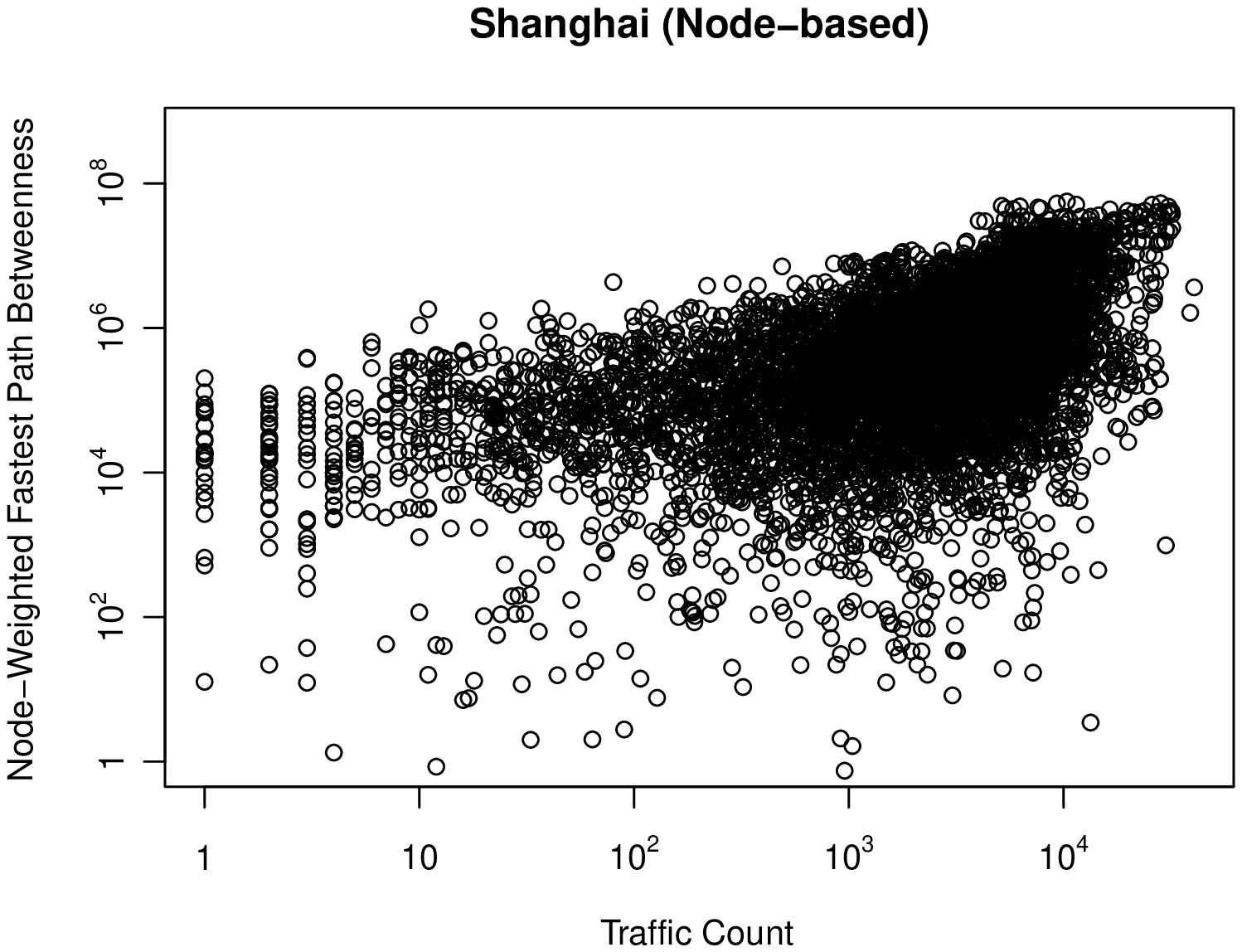}\includegraphics[width=0.49\columnwidth]{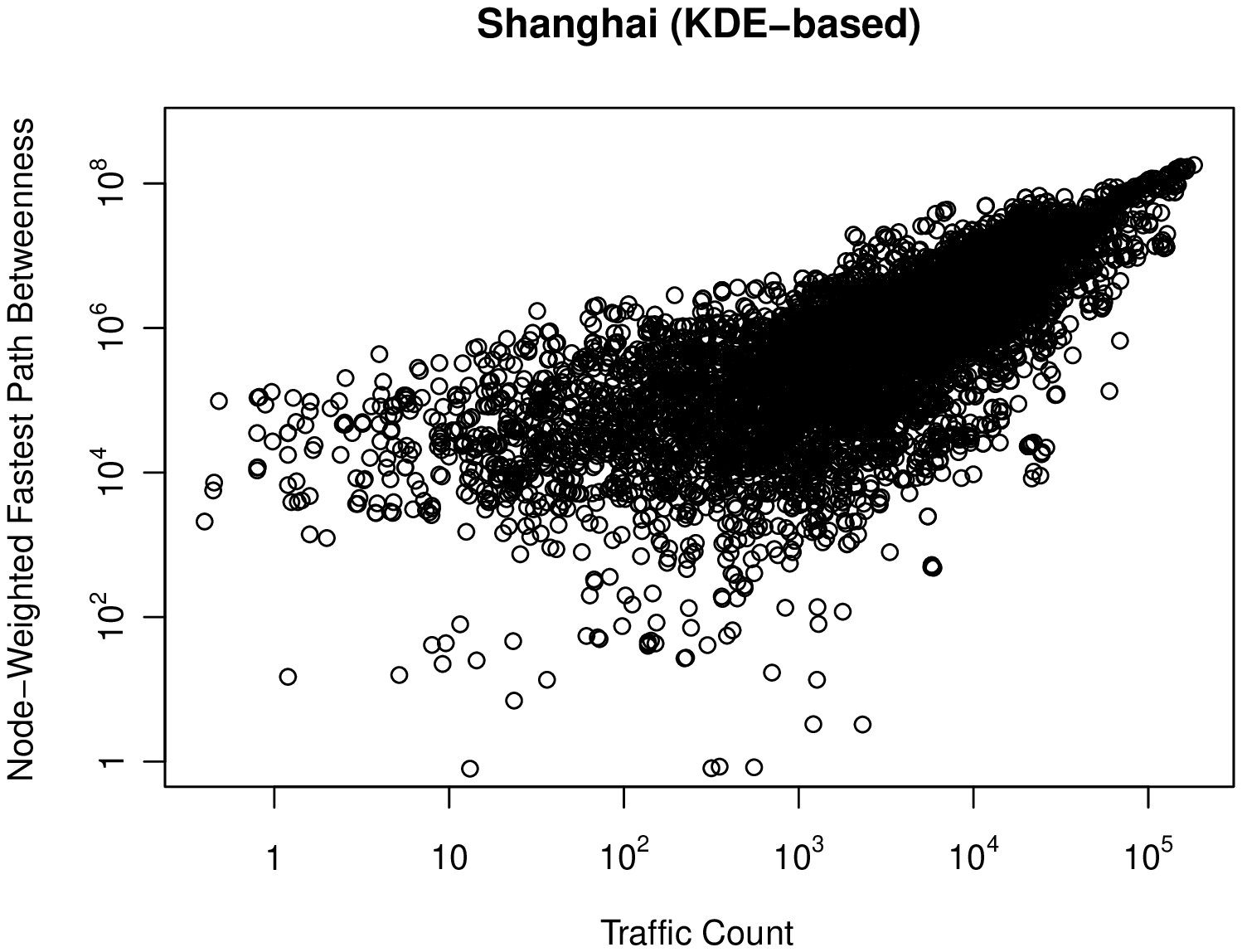}
   \caption{Scatter plots of Node-weighted betweenness against the observed and interpolated traffic count in the two cities. The plots are in log-log scale and clear positive trends can be observed in all cases. The Pearson's $r$ from the correlation studies are reported in Tables \ref{5tab1} and \ref{5tab2}.}
   \label{corr}
\end{figure}

In short, we have demonstrated that in both cities, combining both topological properties and liveliness information greatly improves performance of traffic prediction. Figure \ref{corr} shows the scatter plots of node-weighted betweenness against the routing-service interpolated traffic network in the two cities with the two different correlation methodologies.

\section{Conclusion and Future Work}\label{5conclusions}
In this literature, more than a hundred million GPS taxi traces in two intra-city urban networks of San Francisco and Shanghai have been analysed with a weighted network perspective. We have reviewed relevant literature on spatial network analysis, applications of large-scale human mobility analysis, as well as existing work on topological-based traffic flow analysis. We have discussed methodologies to effectively simplify the urban networks as well as estimate traffic flow from raw data. We have argued that based on pure topological features, like a lot of previous work did, would be insufficient to consistently predict traffic flow due to the inherently flawed assumption which some centrality measures carry. A novel framework which allows node weight and travel speed to be incorporated into traditional betweenness analysis has been shown to greatly improve traffic prediction performance. Based on the topological framework, we have also identified driving behaviours which potentially differ during rush hour and non-rush hour traffic.

Throughout the paper, we have stressed that a node weighting scheme which reflects a junction's capability to generate and attract traffic is vital to the success of traffic flow prediction. It is therefore an important future work to devise a suitable scheme for weighting nodes prior to betweenness analysis. A key drawback for analysis on an undirected network, however, is that not only does it ignore the direction restriction of the roads, it also disregards the turning restrictions at many junctions. The lack of such information translates to the inaccurate shortest-path estimations from different sources to destinations in different directions.

In the previous section, we have also discussed the further need to understand the effect of the characteristic, size or even culture of the cities have on traffic flow analysis. In particular, the definition of an ``intra-city" network, i.e., its size, scale, and granularity, should require centrality measures of different scales to adjust to the fact that an average vehicle would tend to travel within a bounded area of a city. It is also important to understand the potential bias towards nodes located in the center of the network given an intra-city network of a certain size.

We maintain that incorporating the perceived or expected travel time of a road into its weight is essential but note that other factors such as its width, permeability, and charge may be as well important. We foresee that as GPS routing system continues to become standardised equipments on vehicular travel, the notion of bounded rationality in human way-finding becomes less and less relevant. As such, it leads us a potential future work where one can investigate the all-pairs routing paths given by GPS navigation systems rather than fastest paths.

To conclude, we believe that with the incorporation of locale information, more realistic network representation to include travel restrictions, and better understanding of way-finding behaviours from GPS routing services, topological-based traffic flow analysis will be a beneficial and effective tool for urban and infrastructure planning.

\ack
The Shanghai taxi data was obtained from Wireless and Sensor networks Lab (WnSN), Shanghai Jiao Tong University, for which we are grateful. The city graphs used in this manuscript are generated using Tulip 3.5.0~\cite{Auber2003}.

\clearpage
\appendix
\section{Node-weighted Brandes Algorithm}
The detailed implementation of the modified Brandes algorithm which allows weighted edges and nodes is given below in Algorithm \ref{5algor1}.

\SetKwComment{Comment}{/* }{ */}
\begin{algorithm}
\scriptsize
\SetKwInOut{Input}{Input} \SetKwInOut{Output}{Output}
\BlankLine
\Input{$G = (V, E)$,$demand[\ ]$(\emph{Node weight}), $weight(v,v')$ (\emph{Edge weight of (v, v')})}
\Output{$C_B[\ ]$ (\emph{The approximation of node-weighted betweenness of each node})}
\BlankLine
\Begin{
$C_B[v] \leftarrow 0, v \in V$\;
\ForEach{$s \in V$}
{\CommentSty{/*s is the source node in each iteration */}\\
$S \leftarrow$ empty stack\;
\CommentSty{/* P[] is a list of predecessors in the shortest path */}\\
$P[w] \leftarrow$ empty list, $w \in V$\;
\CommentSty{/* counters for number of shortest paths */}\\
$\sigma[t] \leftarrow 0,\ t \in V$; $\sigma[s] \leftarrow 1$\;
\CommentSty{/* distances from source */}\\
$d[t] \leftarrow -1,\ t \in V$; $d[s] \leftarrow 0$;\\
\CommentSty{/* a Priority queue of nodes ordered with increasing d[node] */}\\
$Q \leftarrow $ empty queue; enqueue $s \rightarrow Q$\;
\CommentSty{/* Dijkstra's algorithm, also counting the number of
equal distance shortest paths to reach each node */}\\
\While{$Q$ not empty}
{
dequeue $Q \rightarrow v$\;
push $v \rightarrow S$\;
\ForEach{$neighbour\ w\ of\ v$}
{
\CommentSty{/* w found for the first time? */}\\
\If {$d[w] < 0$}
{ $d[w] \leftarrow d[v]\ +\ weight(v,w)$\;
 enqueue $w \rightarrow Q$\;
}
\CommentSty{/* shorter path to w via v? */}\\
\If {$d[w] \ > d[v]\ +\ weight(v,w)$}
{
$d[w] \leftarrow d[v]\ +\ weight(v,w)$\;
$\sigma[w] \leftarrow \sigma[v]$\;
$P[w] \leftarrow new\ list();$ append $v\ \rightarrow\ P[w]$\;
\CommentSty{/* reorder w in Q with the value of d[w] */}\\
$Q.decreaseKey(w)$\;
}
\ElseIf {$d[w] \ =$ $d[v]\ +\ weight(v,w)$}
{
\CommentSty{/* one of the shortest paths to w via v */}\\
$\sigma[w] \leftarrow \sigma[w]\ +\ \sigma[v]$\;
append $v \rightarrow P[w]$\;
}
}
}
\CommentSty{/* counters for number of shortest paths from s passing through each node*/}\\
$\delta[v] \leftarrow 0,\ v\ \in\ V$\;
\CommentSty{/* S returns vertices in decreasing order of distance from s */}\\
\While{$S$ not empty}
{
 pop $S \rightarrow\ w$\;
 \ForEach {$v \in P[w]$}
 {
\lnlset{5lineA}{a}$\delta[v] \leftarrow \delta[v] + \frac{\sigma[v]}{\sigma[w]} \cdot (demand[s] \cdot demand[w]\ +\ \delta[w])$\;
\If{$w \neq s$}
{$C_B[w] \leftarrow C_B[w]+\delta[w]$\;}
}
}
}
}
\caption{\label{5algor1}Full implementation of the node-weighted betweenness centrality algorithm.}
\end{algorithm}

\clearpage
\section*{References}
\bibliography{ref}

\end{document}